\documentclass[11pt,twoside]{article}
\usepackage[pdftex]{graphicx}
\usepackage[numbers,sort&compress]{natbib}
\usepackage{amsmath}
\usepackage{amssymb}
\usepackage{braket}
\usepackage{bbm}

\usepackage{color}

\begin{document}
\newcommand{\pst}{\hspace*{1.5em}}

\newcommand{\rigmark}{\em Journal of Russian Laser Research}
\newcommand{\lemark}{\em Volume 30, Number 5, 2009}


\newcommand{\ketbra}[2]{\ket{#1}\!\bra{#2}}
\newcommand{\Id}{\mathbbm{1}}

\thispagestyle{plain}

\label{sh}


\begin{center} {\Large \bf
\begin{tabular}{c}
Quantum state identification of qutrits 
\\[-1mm]
via a nonlinear protocol
\end{tabular}
 } \end{center}

\bigskip

\bigskip

\begin{center} {\bf
P.~V.~Pyshkin$^{1}$, A.~G\'abris$^{2,1}$, O.~K\'alm\'an$^{1}$, I.~Jex$^{2}$ and T.~Kiss$^{1*}$
}\end{center}

\medskip

\begin{center}
{\it
$^1$Institute for Solid State Physics and Optics, Wigner Research Centre, Hungarian Academy of Sciences\\
P.O. Box 49, H-1525 Budapest, Hungary

$^2$Czech Technical University in Prague, Faculty of Nuclear Sciences and Physical Engineering \\ 
B\v rehov\'a 7, 115 19 Praha 1, Star\'e M\v esto, Czech Republic.
}
\smallskip

$^*$Corresponding author e-mail: \texttt{kiss.tamas@wigner.mta.hu}\\
\end{center}

\begin{abstract}\noindent
We propose a probabilistic quantum protocol to realize a nonlinear transformation of qutrit states, which by iterative applications on ensembles can be used to distinguish two types of pure states. The protocol involves single-qutrit and two-qutrit unitary operations as well as post-selection according to the results obtained in intermediate measurements. We utilize the nonlinear transformation in an algorithm to identify a quantum state provided it belongs to an arbitrary known finite set. The algorithm is based on dividing the known set of states into two appropriately designed subsets which can be distinguished by the nonlinear protocol. In most cases this is accompanied by the application of some properly defined physical (unitary) operation on the unknown state. Then, by the application of the nonlinear protocol one can decide which of the two subsets the unknown state belongs to thus reducing the number of possible candidates. By iteratively continuing this procedure until a single possible candidate remains, one can identify the unknown state.  
\end{abstract}

\medskip

\noindent{\bf Keywords:}
quantum measurement, quantum control, quantum state identification.

\section{Introduction}
\pst
Measurement on a quantum system inevitably affects its state. One of the questions J\'ozsef Janszky was intrigued by in his last active years was how one can design useful protocols involving post-selection based on measurement results \cite{QuantumScissors, Janszky}. The power of measurement-based protocols can be used in quantum state purification~\cite{Hiromichi, aschauer_multiparticle_2005, cooling_by_feedback_control}, as well as for quantum state engineering~\cite{Piani2014_entanglment_by_measurements, Streltsov2011_entanglment_by_measurements, Wu_entanglement_generation, Pyshkin_compression, Filippov2017}, in particular also to cool down quantum systems to their ground-state~\cite{Li2011, gsc_paper, hertzberg_back-action-evading_2010, rocheleau_preparation_2010}. One can exploit the nonlinear nature of this type of protocols for enhancing initially small differences between quantum states \cite{Gilyen}. 
 
Discrimination of nonorthogonal quantum states is an important task for applications of quantum information and quantum control~\cite{Nielsen2000}. Various protocols have been proposed for efficient quantum state discrimination (QSD) (see reviews~\cite{Barnett09,Kwek2015}). A crucial ingredient of these methods is to have an ensemble of identical quantum systems for implementing QSD~\cite{mack_enhanced_2000, Torres2017, Zhang2018, Kalman2018}. Measurement-induced nonlinear dynamics is experimentally feasible in quantum optics~\cite{Xu2014}, and it has been shown~\cite{Torres2017, Kalman2018} that nonlinear quantum transformations could be a possible way for implementing QSD of two-level quantum systems. In this report we propose a scheme which can be used for QSD of three-level quantum systems.  Such systems are studied as candidates for quantum processing also experimentally, see e.g.~\cite{AbdumalikovJr2013}. 

Quantum state identification (QSI) is a problem where one has to decide whether an unknown quantum state is identical to one of some reference quantum states. In the original formulation of the problem \cite{Hayashi2005, Hayashi2006} the unknown pure state has to be identified with one of two or more reference pure states, some or all of which are unknown, but a certain number of copies of them are available \cite{Herzog2008, Herzog2016}.    

In this paper we design a quantum protocol based on post-selection where the difference between the absolute values of two coefficients in the expansion of the quantum state of a three-level system (qutrit) is enhanced. The protocol is thus capable of decreasing the overlap of initially nonothogonal ensembles of systems, according to a specific property of the states. We show that one can build an algorithm around this protocol which solves a quantum-state-identification type of problem where a finite number of reference states is classically given.   

\section{Nonlinear transformation of qutrit states}
\begin{figure}[t]
	\begin{center}
	\includegraphics{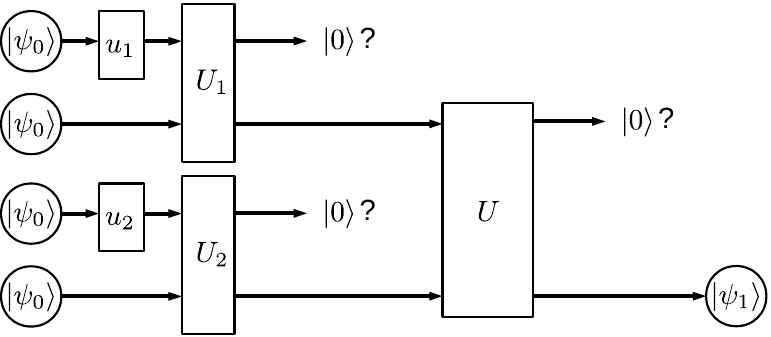} 
	\end{center}
	\caption{Scheme of the non-linear transformation $\ket{\psi_0}\rightarrow\ket{\psi_1}$ as a quantum circuit. Each line represents a qutrit.}
	\label{fig1}
\end{figure}

\pst
We consider an ensemble of identically prepared quantum systems in the state parametrized by two complex parameters, $z_1$ and $z_2$ as
\begin{equation}
\ket{\psi_0} = \mathcal{N}\left( \vphantom{\frac{1}{1}}\ket{0} + z_1\ket{1} + z_2\ket{2}\right),
\label{initial_1}
\end{equation}
with $\mathcal{N} = (1 + |z_1|^2 + |z_2|^2)^{-1/2}$ chosen such that $\| \ket{\psi_0} \|=1$. In the following we describe a protocol that allows us to distinguish between cases: (i) $|z_1|>|z_2|$ and (ii) $|z_1|<|z_2|$, regarding the parametrization. The core of this procedure is the nonlinear transformation schematically depicted on Fig.~\ref{fig1} as a quantum circuit diagram,
with the single-qutrit unitary operators defined as
\begin{equation} \label{single_Us}
\begin{array}{r@{}l}
&u_1 = \ketbra{0}{2} + \ketbra20 + \ketbra{1}{1}, \\
&u_2 = \ketbra{0}{1} + \ketbra{1}{0} + \ketbra{2}{2}, \\ 
\end{array}
\end{equation}
and the two-qutrit operators as
\begin{equation} \label{two_Us}
\begin{array}{r@{}l}
&U_1 = \ketbra{01}{11} + \ketbra{11}{01} + (\Id - \ketbra{01}{01} - \ketbra{11}{11}), \\ 
&U_2 = \ketbra{02}{22} + \ketbra{22}{02} + (\Id - \ketbra{02}{02} - \ketbra{22}{22}), \\
&U = \ketbra{01}{10} + \ketbra{10}{01}  + (\Id - \ketbra{01}{01} - \ketbra{10}{10}).
\end{array}
\end{equation}

The procedure starts by taking two pairs of the system in initial state~$\ket{\psi_0}$. Then one member of each pair is transformed by a single-qutrit unitary $u_j$ ($j=1,2$), after which $U_j$ acts on the pair of qubits as a whole, followed by perfoming selective projective measurements $P = \ketbra{0}{0}$ on the first system of each pair. If both results are ``yes'' then we take the unmeasured systems from each pair and apply a joint unitary operator~$U$ on them. Then we perfom again a projective measurement $P$ on the first system in the pair, and if the result is again ``yes'' then the unmeasured system transforms into the state
\begin{equation}\label{psi1}
\ket{\psi_1} = \mathcal{N}'\left(\vphantom{1^1}\ket{0} + \frac{z_1}{z_2}z_1\ket{1} + \frac{z_2}{z_1}z_2\ket{2}\right).
\end{equation}

By appling the above procedure to the entire ensemble of qutrits (always taking two pairs at a time) we arrive at a new albeit smaller ensemble constituted by identical states. We can interpret this as transforming an ensemble described by the state $\ket{\psi_0}$ to an ensemble described by the state $\ket{\psi_1}$.

\begin{figure}[ht]
	\begin{center}
	\includegraphics[scale=.8]{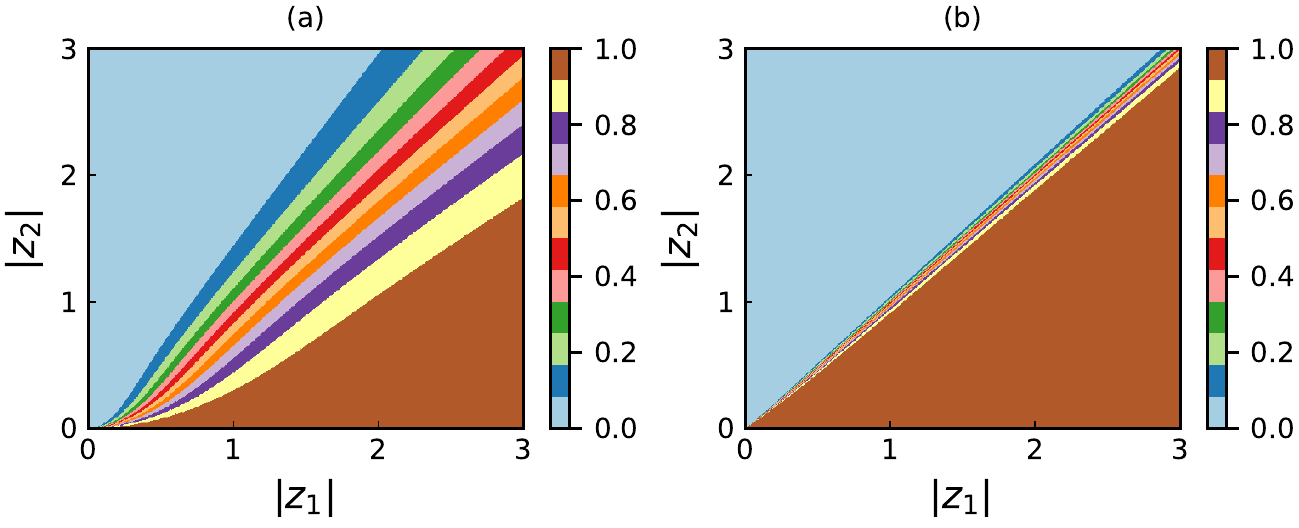}
	\end{center}
	\caption{Probability of obtaining the state $\ket{1}$ in a measurement after $M=1$ iteration (a) and $M=3$ iterations (b) of the nonlinear transformation as a function of the parameters of the initial state given by Eq.~(\ref{initial_1}). The thickness of the transient boundary is indicative of the error margin of correct identification.
	}
	\label{fig2}
\end{figure}

The transformation $\ket{\psi_0} \to \ket{\psi_1}$ is a nonlinear vector mapping: 
\begin{equation} \label{mapping}
\vec{f}^{(n)} = \{f_1^{(n)}, f_2^{(n)}\}, \quad \vec{f}^{(n)} = \vec{f}(\vec{f}^{(n-1)}), \quad \vec{f}^{(0)} = \{z_1, z_2\},
\end{equation}
where $f_1^{(n)} = {f_1^{(n-1)}}^2 / f_2^{(n-1)}$ and $f_2^{(n)} = {f_2^{(n-1)}}^2 / f_1^{(n-1)}$. Thus, the result of~$M$ iterations will be state~$\ket{\psi_M} \propto \ket{0}+f_1^{(M)}\ket{1} + f_2^{(M)}\ket{2}$. The map~(\ref{mapping}) has two attractors: $\{\infty, 0\}$ and $\{0, \infty\}$. Therefore, if $|z_1|\neq|z_2|$ we will have for some relatively large~$M$: $\ket{\psi_M}\approx \ket{1}$ in the case of $|z_1| > |z_2|$, and $\ket{\psi_M}\approx \ket{2}$ in the case of $|z_1| < |z_2|$. We can distinguish these two cases, up to a certain error margin, by performing the projective measurement~$\ket{1}\bra{1}$ on the system in the state~$\ket{\psi_M}$, which allows us to draw conclusions regarding initial state $\ket{\psi_0}$. In Figs.~\ref{fig2}(a) and \ref{fig2}(b) we show the probability of obtaining the state $\ket{1}$ in a measurement after one iteration ($M=1$) and three iterations ($M=3$) of the nonlinear transformation of Eq.~(\ref{psi1}), respectively. The border between regions with high and low probability corresponds to the $|z_1| = |z_2|$ condition, and this border becomes sharper with increasing $M$. Thus, the reliablity of QSI increases with increasing $M$. Note, that the above discussed probability describes the precision of the discrimination process at the end of the protocol, yielding an {\em error margin} on making the right conclusion about the given initial state \cite{hayashi_state_2008}. Another relevant quantity is the {\em survival probability} which describes the post-selection process that is based on the intermediate projective measurements shown in Fig.~\ref{fig1}. We will discuss this probability at the end of this Section. 

\begin{figure}[ht]
	\begin{center}
	\includegraphics[scale=.85]{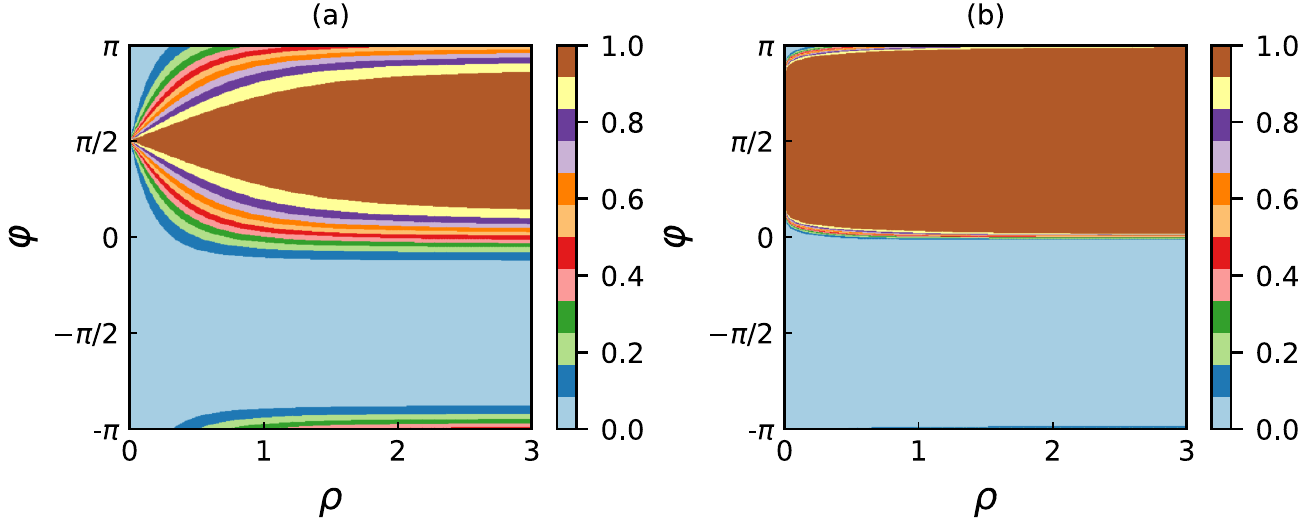}
	\end{center}
	\caption{Probability of measuring state $\ket{1}$ after the application of the single-qutrit unitary operation of Eq.~(\ref{Rot}) and then the subsequent iteration ($M=1$ for (a), and $M=3$ for (b)) of the nonlinear transformation as a function of the parameters of the initial state of Eq.~(\ref{initial_2}). }
	\label{fig3}
\end{figure}

The nonlinear transformation of Eq.~(\ref{psi1}) itself cannot distinguish two states in which $|z_1|=|z_2|$, as can be seen from Fig.~\ref{fig2}. However, a properly chosen single-qutrit unitary operation can be used to make the magnitudes of these coefficients different so that the subsequent nonlinear transformation can distinguish such states. In order to show this, let us consider input states of the form  
\begin{equation}
\ket{\psi_0} = \mathcal{N}(\ket{0} + \rho\ket{1} + \rho \exp(i\varphi)\ket{2}),
\label{initial_2}
\end{equation}
where $\rho,\varphi\in\mathbb{R}$. Then, apply the following single-qutrit unitary transformation~$R$
\begin{equation} \label{Rot}
R =  \left(\begin{array}{ccc}
1  & 0 & 0  \\
0  & i/\sqrt2 & 1/\sqrt2  \\
0  & -1/\sqrt2 & -i/\sqrt2 
\end{array} \right)  
\end{equation}
on each initial state. Due to this operation the transformed state will be 
\begin{equation}
R\ket{\psi_0} = \mathcal{N}(\ket{0} + z_1'\ket{1} + z_2'\ket{2}),
\label{Rpsi0}
\end{equation}
where
\begin{equation}
|z_1'|= \rho\sqrt{1 + \sin\varphi}, \quad |z_2'|= \rho\sqrt{1 - \sin\varphi}. 
\end{equation}
Therefore, we can treat the problem as before, since when $\varphi\neq0$ then $|z_1'|\neq|z_2'|$ and applying the nonlinear transformation on the state of Eq.~(\ref{Rpsi0}) we can distinguish the two different situations: $0<\varphi<\pi $ (corresponding to $|z_1'|>|z_2'|$), and $-\pi<\varphi<0 $ (corresponding to $|z_1'|<|z_2'|$). In Figs.~\ref{fig3}(a) and \ref{fig3}(b) we show the probability of measuring state~$\ket{1}$ as a function of the initial values of $\rho$ and $\varphi$.

\begin{figure}
	\begin{center}
	\includegraphics[width=12cm]{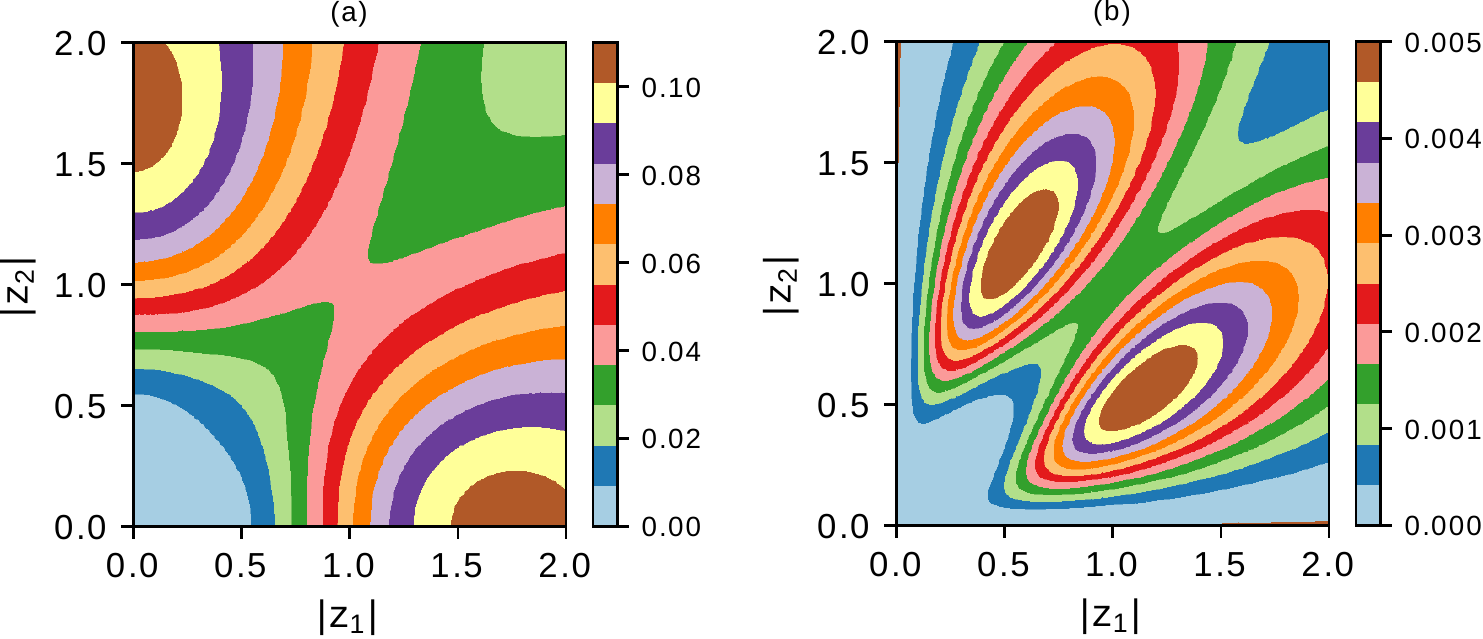}
	\end{center}
	\caption{The overall survival probability of the nonlinear protocol for $M=1$ (a), and $M=2$ iterations (b) as a function of the initial parameters. Note that in the latter case the survival probability is given by the product of the $P_{s}$'s of the two steps. To determine the $P_{s}$ of the second step, the transformed values of $z_{1}$ and $z_{2}$ were substituted into Eq.~(\ref{Ps}).}
	\label{fig4}
\end{figure}

In practical situations the necessary number of iterations of the nonlinear protocol is determined by the probabilities depicted in Fig.~\ref{fig2} and \ref{fig3}. However, due to the fact that our nonlinear protocol is based on selective measurements, we need to have a relatively large number of identical qutrits in the initial state~$\ket{\psi_0}$. This can be characterized by the survival probability in a single step 
\begin{equation}\label{Ps}
P_s = \frac{ |z_1|^2|z_2|^2 + |z_1|^6 + |z_2|^6  }{( 1 +|z_1|^2 + |z_2|^2   )^4},
\end{equation}
which is the product of the three probabilities to measure $\ket{0}$ (see Fig.~\ref{fig1}).
From Eq.~(\ref{Ps}) it can be seen that the survival probability is very small (i.e., the protocol is very source demanding) when both $z_1, z_2 \rightarrow 0$ or $z_{1(2)}\rightarrow\infty$. However, in cases when only a few iterational steps are expected to give a conclusive answer for discrimination, then the survival probability is also realtively high (see Fig.~\ref{fig4}). 

\section{Quantum state identification of qutrits}
\pst
Let us assume we have a ``black box'' that produces qutrits in the quantum state ~$\ket{\psi_{?}}$, which is unknown to us. What we know is that it belongs to a {\em finite set} $S$ of possible qutrit states which we denote by
\begin{equation}
  \label{K-set}
  S = \left\{ \ket{\psi_i} = \mathcal{N}_i\left(\vphantom{1^1}  \ket{0} + z_{i1}\ket{1} + z_{i2}\ket{2}\right) \mathop{\vert} i = 1, \ldots, K \right\},
\end{equation}
with $z_{ij} \in \mathbbm{C}$ being complex numbers. We also introduce the function $N\colon  S \rightarrow N(S)$ denoting the number of elements in the set $S$, yielding $N(S)=K$ in the particular case. The problem of quantum state identification~(QSI) is to find $w$ such that $\ket{\psi_{?}} = \ket{\psi_w}$ by applying quantum operations on the ensemble produced by the black box. 
For the sake of simplicity let us assume that $|z_{i1}|\neq|z_{i2}|$ for all $i\in(1,2,\dots, K)$. In case we have $|z_{n1}|=|z_{n2}|$ for some $n$ we can apply a unitary rotation similar to the one in Eq.~(\ref{Rot}) to transform the coefficients into new ones of unequal magnitude.

\begin{figure}
	\begin{center}
	\includegraphics[width=8.6cm]{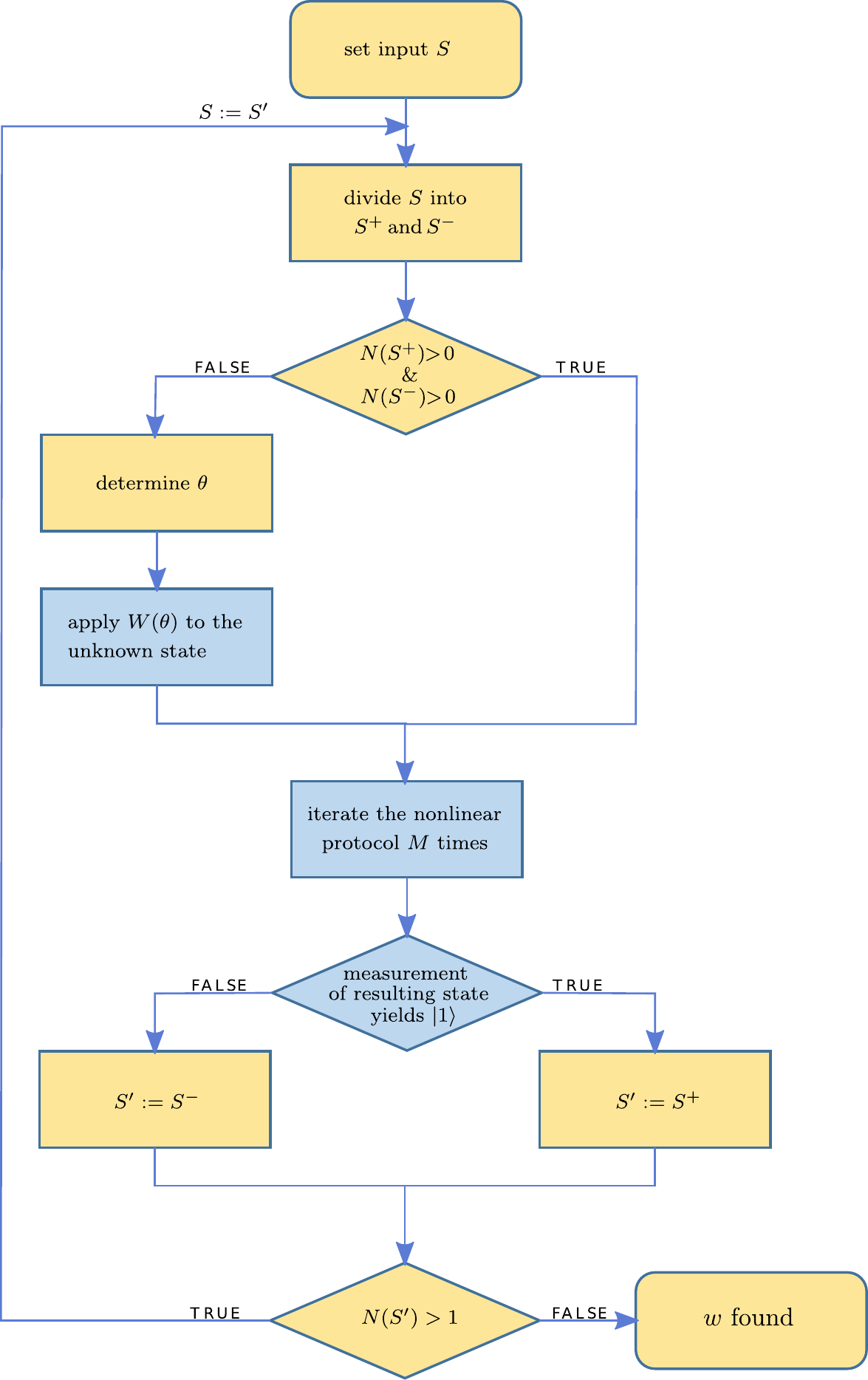}
	\end{center}
	\caption{The flow chart of the proposed QSI algorithm employing the non-linear transformation described by the map in Eq.~(\ref{mapping}). }
	\label{fig5}
\end{figure}

Before discussing the QSI algorithm itself, let us consider the following unitary rotation:
\begin{equation} \label{W}
W(\theta) = \left( \begin{array}{ccc}
1 &  0 & 0  \\
0 & \cos\theta & \sin\theta  \\
0 & - \sin\theta & \cos\theta 
\end{array} \right).
\end{equation}
This can transform a quantum state of the form of Eq.~(\ref{initial_1}) with $|z_{1}|\neq|z_{2}|$ into the state with $|z_{1}|=|z_{2}|$ if $\theta$ is chosen in the following way:
\begin{equation}\label{teta_1}
\tan2\theta = \frac{|z_2|^2 - |z_1|^2}{z_1z_2^* + z_1^*z_2}.
\end{equation}
Moreover, by choosing a different angle $\theta'$ in such a way that $\theta < \theta' < \pi/4$ (for $\theta>0$) or $-\pi/4 < \theta' < \theta$ (for $\theta<0$) we can change the sign of $|z_{1}| - |z_{2}|$. We take advantage of such transformations in our quantum state identification algorithm, which is depicted in Fig.~\ref{fig5} by a flowchart. 

The QSI algorithm of Fig.~\ref{fig5} is composed of calculational steps (with yellow background) in which the parameters of the necessary physical operations are determined and these are followed by the actual physical operations on the qutrit ensemble (with blue background). The description of the algorithm is the following
\begin{description}
\item[Inputs:] Set $S$ as given by Eq.~(\ref{K-set}) with $K\geq 2$ elements. Ensemble of qutrits in an unknown state $\ket{\psi_{?}} \in S$.
\item[Output:] Number $w$ such that $\ket{\psi_w} = \ket{\psi_{?}}$ up to a desired error margin.
\item[Procedure:] ~\par
  \begin{enumerate}\itemsep1pt
  \item Divide $S$ into $S^+$ (containing states with $|z_{n1}|>|z_{n2}|$) and $S^-$ (containing states with $|z_{m1}|<|z_{m2}|$).
  \item If both $N(S^{+})>0$ and $N(S^{-})>0$ then skip to \textbf{step \ref{step:5}}, otherwise continue to the next step.
  \item Determine a proper~$\theta$ with which $S$ can be divided into subsets $S^+$ and $S^-$ with $N(S^{\pm})>0$.
  \item Apply the unitary rotation $W(\theta)$ of Eq.~(\ref{W}) to the ensemble with the unknown state $\ket{\psi_{?}}$.
  \item\label{step:5} Apply iteratively $M$ times the nonlinear protocol of Fig.~\ref{fig1} to a sufficiently large ensemble representing the unknown state $\ket{\psi_{?}}$. 
  \item Make a projective measurement to decide whether the unknown state belongs to the set $S^{+}$ or to $S^{-}$, i.e.\ whether $|z_{w1}|>|z_{w2}|$ or $|z_{w1}|<|z_{w2}|$ should be satisfied by $w$.
  \item If the result is ``true'' (i.e., in the projective measurement the state $\ket{1}$ was found) then we define $S'$ to be equal to $S^+$, if the result is ``false'' (i.e., in the projective measurement the state $\ket{2}$ was found) then we set $S'$ to $S^-$.
  \item If the number of elements of the new set $N(S')=1$ then the single element of the set is equal to the unknown quantum (apart from possible $W(\theta)$ rotations), thus $w$ has been found. If $N(S')>1$ then repeat the whole procedure from \textbf{step 1} with $S$ set to $S'$. 
  \end{enumerate}
\end{description}
  


\begin{figure}[h]
	\begin{center}
	\includegraphics[width=8.0cm]{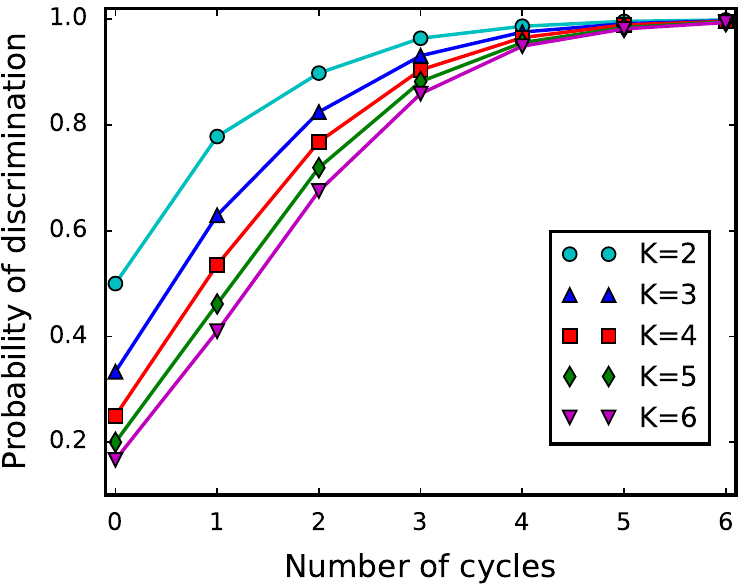}
	\end{center}
	\caption{The average probability of correctly identifying the unknown quantum state among the given set (of size $K$) of qutrit states as a function of the number of iterations~$M$ of the nonlinear transformation in step \ref{step:5}. No iterations yield the probability $1/K$.}
	\label{fig6}
\end{figure}

The efficiency of our algorithm is illustrated in Fig.~\ref{fig6}. We numerically simulated the QSI algorithm of Fig.~\ref{fig5} by choosing $20000$ random realizations of sets of qutrit states of different sizes from $K=2$ to $K=6$. The coefficients $z_{ij} = \rho_{ij}\exp( i \varphi_{ij})$ of the states were given by choosing $\rho_{ij}$ and $\varphi_{ij}$ randomly from a uniform distribution in the interval~$ [1/2,2)$ and $ [0,2\pi)$, respectively. The integer parameter $w$ was randomly chosen in the range $1,2,\dots, K$. As can be seen from Fig.~\ref{fig6} the results of our algorithm scale well with increasing set size. In fact, it is comparable to the well-known ``weighing puzzle``, in which someone has to find a ``false coin'' in a given set of coins by using balanced scales \cite{Chudnov2015}. In our numerical simulations if one of the subsets $S^+$ or $S^-$ was empty, then, for the states of the non-empty subset we calculated the ordered set $\theta_1<\theta_2<\dots<\theta_D$ by using Eq.~(\ref{teta_1}) (where $D$ is the size of non-empty subset). We counted the number $n^{+}$ ($n^{-}$) of positive (negative) $\theta_{i}$'s  and calculated their difference $d=n^{+}-n^{-}$. Then, we determined $\theta$ according to Table~\ref{td}. Let us note that if $\theta_l = \theta_m$ for some $1\leq l,m \leq K$, then this procedure fails. As can be seen from Eq.~(\ref{teta_1}) this occurs when $z_{mj} = Az_{lj}$, where $j=1,2$, and $A$ is some constant. In order to solve this problem we can apply the $u_2$ single-qutrit unitary transformation of Eq.~(\ref{single_Us}) in step 3 to every member of the set~(\ref{K-set}). Thus, if we initially have two states $\ket{\psi_l} = \mathcal{N}_l (\ket{0} + z_{l1}\ket{1}+z_{l2}\ket{2})$ and $\ket{\psi_m} = \mathcal{N}_m (\ket{0} + Az_{l1}\ket{1}+Az_{l2}\ket{2})$, then after this transformation we will have $\ket{\psi_l\rq} = \mathcal{N}_l' (\ket{0} + \frac{1}{z_{l1}}\ket{1}+\frac{z_{l2}}{z_{l1}}\ket{2})$ and $\ket{\psi_m\rq} = \mathcal{N}_m' (\ket{0} + \frac{1}{Az_{l1}}\ket{1}+\frac{z_{l2}}{z_{l1}}\ket{2})$ and the corresponding $\theta$ angles will be different. 

\begin{table}[]
	\centering
	\begin{tabular}{|c|l|}
		\hline
		$d=n^{+}-n^{-}$ & optimized $\theta$ \\ \hline
		$d=-D$ & $\theta = (\theta_{\lfloor D/2\rfloor} + \theta_{\lfloor D/2\rfloor+1})/2$ \\ \hline
		$-D<d\leq-2$ &  $\theta = (\theta_{\lfloor - d/2 \rfloor} + \theta_{\lfloor -d/2\rfloor+1})/2$ \\ \hline		
		$-1\leq d \leq 1$ &   $\theta = (\theta_1 - \pi/4)/2$ \\ \hline	
		$2\leq d < D$ &   $\theta = (\theta_{\lfloor D - d/2 \rfloor} + \theta_{\lfloor D-d/2\rfloor+1})/2$ \\ \hline	
		$d=D$ & $\theta = (\theta_{\lfloor D/2\rfloor} + \theta_{\lfloor D/2\rfloor+1})/2$ \\ \hline
	\end{tabular}
	\caption{Optimized determination of $\theta$ to divide $S$ into $S^{+}$ and $S^{-}$. We have denoted the lower integer part of a real number $x$ by $\lfloor x \rfloor$.}
	\label{td}
\end{table}

Due to the optimization we apply during every loop when dividing the set $S$, the average number of loops that are needed for the QSI algorithm to complete scales as $\ln K$. In Table~\ref{t1} we show the numerical results with the average number of loops and their standard deviations based on numerical simulations with~$20000$ random realizations of the set given by Eq.~(\ref{K-set}).

\begin{table}[h]
	\centering
	\begin{tabular}{|l|l|l|l|l|l|}
		\hline
		Set size & $K=2$ & $K=3$ & $K=4$ & $K=5$ & $K=6$ \\ \hline
		Average number of loops & $1$ & $1.89$ & $2.58$ & $3.16$ & $3.66$ \\ \hline
		Standard deviation & $0$ & $0.31$ & $0.49$ & $0.57$ & $0.63$ \\ \hline		
	\end{tabular}
	\caption{Average number of loops of the QSI algorithm and their standard deviations for different set sizes.}
	\label{t1}
\end{table}




\section{Summary}
\pst
We have presented a probabilistic scheme to realize a nonlinear transformation of qutrit states with two stable attracting fixed points. The nonlinear transformation is defined on an ensemble of quantum systems in identical quantum state, and in each elementary operation it uses two pairs of systems to probabilistically produce one system in the transformed state. Therefore, the protocol requires at least exponential resources in terms of the size of the ensemble as a function of the number of iterations. 

The nonlinear map can be used to find out whether a given unknown pure state belongs to the subset converging to one of the attractive fixed points. We employed this property for QSI by proposing an algorithm that can be used to identify the quantum state from a finite set $S$ of candidates, with an error margin. We have shown that the number of loops the algorithm uses scales logarithmically with the number of elements $K$ in the set $S$. While the error margin can be made arbitrarily small by increasing the number of iterations $M$ of the nonlinear map, the role and impact of the survival probability remains an open question that deserves further studies.

Our results indicate that probabilistic nonlinear schemes may offer a consistent approach towards QSI of higher dimensional systems, the present study on qutrits being the first step towards this direction.

\section*{Acknowledgments}
\pst
The authors P.~V.~P., O.~K.\ and T.~K.\ were supported by the National Research, Development and Innovation Office (Project Nos.\ K115624, K124351, PD120975, 2017-1.2.1-NKP-2017-00001). In addition, O.~K.\ by the J.~Bolyai Research Scholarship, and the Lend\"ulet Program of the HAS (project No.~LP2011-016). I.~J.\ and A.~G.\ have been partially supported by M{\v S}MT RVO 68407700, the Czech Science Foundation (GA{\v C}R) under project number 17-00844S, and by the project ``Centre for Advanced Applied Sciences,'' registry No.\ CZ.02.1.01/0.0/0.0/16\_019/0000778, supported by the Operational Programme Research, Development and Education, co-financed by the European Structural and Investment Funds and the state budget of the Czech Republic.


\end{document}